# QUANTITATIVE THEORY OF MONEY OR PRICES? A HISTORICAL, THEORETICAL, AND ECONOMETRIC ANALYSIS

José Mauricio Gómez Julián

josemgomez@protonmail.ch, jose.gomezjulian@ucr.ac.cr

**Abstract**

This research studies the relation between money and prices and its practical implications analyzing quarterly data from United States (1959-2022), Canada (1961-2022), United Kingdom (1986-2022), and Brazil (1996-2022). The historical, logical, and econometric consistency of the logical core of the two main theories of money is analyzed using objective bayesian and frequentist machine learning models, bayesian regularized artificial neural networks, and ensemble learning. It is concluded that money is not neutral at any time horizon and that, despite money is ultimately subordinated to prices, there is a reciprocal influence over time between money and prices which constitute a complex system. Non-neutrality is transmitted through aggregate demand and is based on the exchange value of money as a monetary unit.

**Keywords:** Quantitative theory of money, Marx's theory of money, bayesian statistics, machine learning, deep learning, ensemble learning.

**JEL Classification:** C11, C14, E11, E13, E17.

## 1. Introduction

The analysis of the neoclassical theory of money must begin by examining David Hume's theory of money, as pointed out by Schabas & Wennerlind (2011, p. 218), as Hume's essays on money and trade have been relevant for various variants of neoclassical theory such as Friedman (1975), Samuelson (1980), Lucas (1996), among others. For Friedman, in fact, Hume's theory is the starting point for monetary theory, while for Lucas, it marks the beginning of modern monetary theory.

The study of Hume's theory of money implies, as the first step in any study of an economic theory, referring to the historical context in which his theory emerged. Specifically, in this case, it aims to determine whether there was a sufficient quality and quantity of data when the theory under study was formulated and whether there were periods of quantitative stability in the variables under examination.

As verified in (Hume, 1994, pp. 134-135), when Hume formulated the quantity theory of money, he only examined periods when there were revolutions in the value and price of precious metals due to the discovery of American mines and the increase in slave labor (Native Americans), which lowered extraction costs. Additionally, there were gaps in official data related to the expansion and contraction of the circulation (Marx, 2008, pp. 152-157). For these reasons, different classical economists meticulously criticized this position (Steuart, 1767, pp. 394-414), or simply held the opposite view, as in the case of Smith (Curott, 2017, p. 326). The alternative viewpoint proposed by Marx is referred to here as the quantitative theory of prices.

Even Hume himself (Hume, 1994, p. 133) acknowledges that the leveling of goods and gold and silver coins occurs gradually, which is not consistent in rigor with his central thesis of automatic leveling. This gradual leveling will be what opens the possibility for Friedman, Lucas, and others to posit the non-neutrality of money in the short term while maintaining the relationship theorized by Hume between prices and the money supply.

At first glance, the discussion of the relationship between the quantity of money in circulation and nominal GDP may seem irrelevant in terms of the way monetary policy is conducted today since the variable controlled by the monetary authority is the interest rate. This affects the demand for money, and ultimately, the demand for money affects prices. Thus, M1 plays no significant role in price determination.

However, the previous perspective has three drawbacks. The first is that, under neoclassical logic, the demand for money is related to the quantity of money in circulation through the transactional component of that demand. If the quantity of money in circulation increases, *ceteris paribus*, the demand for money also tends to increase, as the increase in M1 leads to higher prices. Therefore, economic agents will need more money for their daily transactions. The second drawback is that the velocity of money circulation is affected by changes in the interest rate, and this velocity affects the quantity of money in circulation (according to the quantity theory of money equation).

Finally, the third drawback is that the discussion about the subordinate variable in the relationship between M1 and prices is ultimately a discussion about the subordinate relationship that exists between variables in the sphere of production and those in the sphere of circulation[1]. This discussion fundamentally revolves around what prices are and, consequently, whether the source of value lies in production or circulation, *i.e.*, the objective theory of value (classical economists and Marx) versus the subjective theory of value (marginalists and neoclassicals).

**2. Validity of the Logical Structure of Cause-Effect Relationships in Hume's Theory of Money**

In *Contribution to the Critique of Political Economy*, Marx conducts an exhaustive critique of the logical structure of cause-effect relationships in Hume's quantity theory of money. This section will present the central aspects of this critique and elaborate on them when necessary.

The first central aspect of Marx's critique concerns the subordination of money to exchange values, which he refers to as the microeconomic and microsocial expression of the subordination of the sphere of circulation to the sphere of production—a macroeconomic and macrosocial phenomenon (Marx, 2008, pp. 149-152).

In the second case, Marx argues that the quantity of value symbols (*i.e.,* money or means of payment available) circulating for a given economy must maintain a certain equilibrium with the quantity of goods and services available for sale. If the quantity of value symbols in circulation is insufficient for the available goods and services, it can lead to a scarcity of means of payment and difficulties in conducting commercial transactions. On the other hand, if there are too many value symbols in circulation, it can lead to price inflation since there are too many means of payment available in relation to the quantity of goods and services.

---

[1] This can be formulated, at the risk of oversimplifying, in the following terms: Does a variable $X$ become subordinate to $Y$ to the extent that $X$ moves away from the sphere of production while $Y$ approaches the sphere of circulation, or does the opposite occur?

Marx suggests that if the quantity of value symbols in circulation decreases or increases below or above its necessary level, there will be a coercive correction in M1 through price adjustments of commodities. This occurs because commodity prices determine the money supply, so adjustments in commodity prices lead to corrections in the money supply. In the base case, this is evident in the fact that a shortage of payment means implies that buyers will have to compete for available goods and services, leading to price increases. Conversely, if there is an excess of means of payment, sellers can raise prices to take advantage of the increased money supply in the market.

Both cases presented by Marx in the previous reference explain how the effects of money non-neutrality are transmitted to prices (both in the short and long term) and the reasons behind this. Thus, non-neutrality is not caused by the money supply determining prices but by the mediate relationship, as pointed out by Marx, between prices and the objective and material foundation of money (gold and silver at that time), which determines the quantity of money in circulation. This implies that this relationship is time-feedback.

The second central aspect of Marx's critique is an epistemological critique of the reasoning that led Hume to develop his theory of money (Marx, 2008, p. 152). This critique consists of the assertion that in the short term, when the production cost of gold and silver decreases, the price of commodities increases. However, this increase only affects the prices of exported commodities since they are the ones exchanged for gold and silver, not as a means of circulation but as commodities. This happens because the reduction in their value is immediately reflected in the supply of gold and silver commodities, whereas the impact of this reduction takes longer to affect gold as a means of payment. It should be noted that in the time when Marx wrote this, most countries used gold and silver as the international exchange medium.

In the international trade of that time, countries that produced certain highly demanded commodities in other countries would export them and receive gold and silver in return. For example, if a country produced a large quantity of silk that was in high demand in other countries, merchants from that country could export silk and receive gold and silver in exchange. In this context, the increase in the supply of gold and silver, because of their re-duced value, had the mentioned impact on the prices of these exported commodities exchanged for gold and silver.

The third central aspect of Marx's critique is a critique of the logical relationship that Hume establishes between the quantity of digits used and the numerical magnitudes within a given numerical system, the incorrect identification of accounting money as a means of circulation and vice versa, and the failure to consider historical events of his time that indicated the need to take into account the exchange value of gold and silver when determining the link between money and prices (Marx, 2008, pp. 154-155).

The fourth central aspect of Marx's critique is a critique of the corollaries that Hume derives from his empirical observation of the statistical behavior of commodity and money prices and how these corollaries form the fragile foundations of his theory (Marx, 2008, pp. 155-157). This can be summarized in the following two points:

1. If metallic currency is a symbol of value, the sum of commodity prices will depend on the quantity of circulating money. Conversely, if the monetary unit is a symbol of value, the quantity of circulating money will depend on the sum of commodity prices. Marx argues that the monetary unit is a symbol of value. This will be studied econometrically in Section V.
2. If money derives its value from prices, there can be more circulating money than the sum of commodity prices. But if money determines prices, there cannot be more money circulating than prices. Marx points out that there can be more money in circulation than the sum of commodity prices.

The confirmation of whether it is metallic currency or the monetary unit that serves as the sign of value cannot be verified directly but indirectly. This will be done through pairwise comparisons of the robustness of the direction of the functional relationship between prices, gold, and circulant currency under different functional forms (linear, quadratic, cubic and quadratic-cubic) to determine, along with the logic of historical evidence, the true direction of the functional relationship between the nominal of variables studied. This will be carried out in Section V.

Regarding point 2, circulating money is defined according to International Monetary Fund (2016, 193) as the monetary aggregate M1, while the sum of commodity prices is, by definition, nominal GDP. If one studies the statistical systems of, for example, the United States, the United Kingdom, Canada, and Brazil, it can be verified that in different years, M1 exceeds nominal GDP. This serves as evidence in favor of Marx's argument. Checking this assertion for the case of Brazil shows that there is no selection bias in the sense that these economies have a higher money supply than the current value of their production because of their currencies being used as reserves by other countries, as an international means of payment, and as currency for loans, among other possible factors. Additionally, Gómez Julián (2016, p. 27) confirmed that M1 for the case of El Salvador is always lower than nominal GDP (although for a shorter analysis period and with different periodicity), which seems to contradict the logic of the previously mentioned selection bias unless it is specified how some factor (such as dollarization) explains this apparent discrepancy.

Marx's central thesis is that the value of money depends on the purchasing power of the commodity or commodities that underlie it, and this purchasing power, in turn, depends on the general level of prices. Such prices, the market prices, are determined by the capitalist competition (which involves the interaction between supply and demand expressed in its essence as technological competition and capital exodus from one economic activity to another seeking to maximize profit), and are going to oscillate around the socially necessary simple average labor required to produce them, which is their center of gravity. The theoretical averages of these market prices are what we call in Marxist economic theory as *prices of production*.

Of course, the above statements are subject to the validity or falsity of the labor theory of value (LTV) as a scientific explanation of price formation since Marx explains exchange values through his LTV.

Regarding the neoclassical theory of value, it is known, following the Cambridge Capital Controversy, that its foundations (the so-called "neoclassical parables") do not provide a scientific explanation of economic (Nitzan & Bichler, 2009, p. 78), (Jiménez, 2011, pp. 213-215), (Pasinetti, 1984, pp. 128, 134, 221), (Samuelson, 1966, pp. 582-583). This is compounded by the aggregation problems exhibited by this theory, which in the most com-plex cases can only be resolved by adopting the labor theory of value from the classical and Marxian traditions (Samuelson, 1962, pp. 194, 203)[2]. This is consistent with the fact that significant databases, such as the Penn World Tables (PWT), do not use marginal productivity of capital as capital remuneration but instead use the real average internal rate of return (TIR) due to the aggregation issue involving different types of capital (Groningen Growth and Development Centre, 2023, pp. 6-7). Other scholars, such as Robinson (1981, p. 166), have recognized that capital can be nothing more than accumulated past labor.

Regarding the PWT, it should be noted that the TIR and the rate of profit calculated "in the Marxist way" (TMG, hereafter) exhibit the same trend-cycle (Wells, 2007, p. 27) and, when considering all economic sectors, converge in distribution.

---

[2] However, he used to criticize it by arguing that "If $Q$ is not a unique product or a fixed-composition bundle of goods, relative price relationships generally will change as real wage rates and profit rates change" (Samuelson, Parable and Realism in Capital Theory: The Surrogate Production Function 1962, 203). Nevertheless, from a Marxist perspective, it is entirely natural for relative prices to change as income distribution changes due to the political and technological characteristics inherent in economic dynamics.

**Figure 1**

*Convergence in Distribution of TIR and TMG*

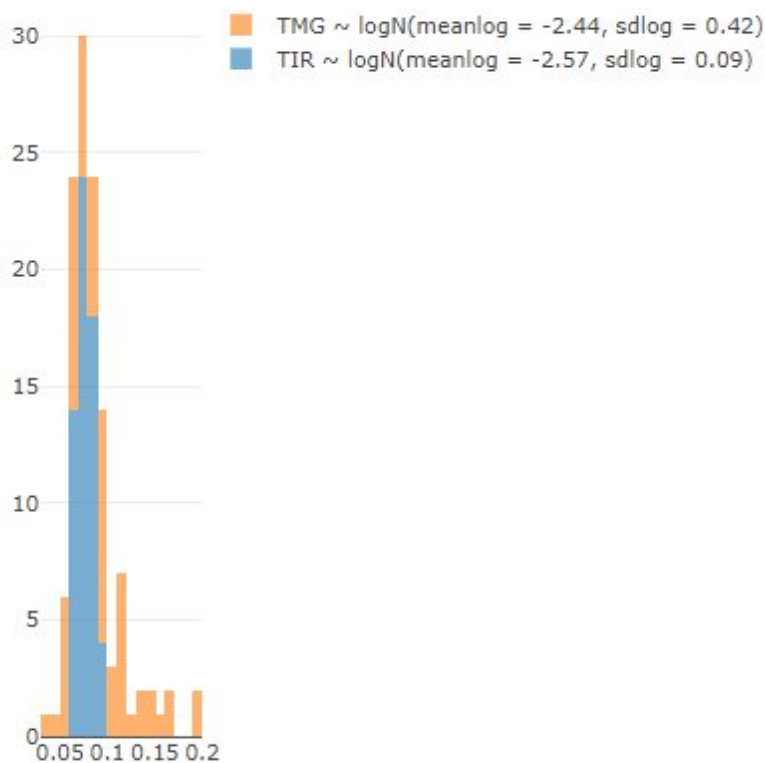

*Source:* Self-made in RStudio, based on the results of empirical distribution fitting by maximum goodness-of-fit and the Bayesian Information Criterion (BIC).

Finally, concerning the so-called "transformation problem", assuming that inventory valuation is done at the cost of reproduction[3] (Potts, 2011, pp. 115-116), then the constant capital (and consequently, also the consumed constant capital and inputs) is a data inherited from the past. As a result, the system has the same number of equations (the equation for the production of Sector I or the producer of means of production, the equation for the production of Sector II or the producer of means of consumption, and the equation for the prices of production) as unknowns (variable capital, the rate of profit, and the prices of production), and it has a unique solution, given the degree of exploitation of labor power[4] (Freeman 1996, pp. 4-7).

---

[3] Valuing at the cost of reproduction (where this cost considers the current technological state and the supply of technologically inferior versions of that inventory) implies the existence of time, and it means that the output prices at the end of the initial production period are equal to the input prices of the following period, and so on.

## 3. Historical Validity of Hume's Theory of Money

### 3.1. Recent Historical Behavior of Prices and Money

The next step is to study the recent historical behavior of prices and the money supply considering the quantity theory of money. The quantity theory of money evolved over time to the version developed by Milton Friedman. How does this theory, in its Friedmanian version, account for real-world events? As noted by Krugman (2008), Milton Friedman and Anna Schwartz argued in *A Monetary History of the United States* that "(...) the Federal Reserve could have prevented the Great Depression, a claim that later, in popular writings, including those by Friedman himself, turned into the assertion that the FED caused the Depression."[5]

At the time, what the FED controlled was the monetary base, and this base increased during the Great Depression (Federal Reserve Bank of St. Louis, 2023). Therefore, it is difficult to argue that the FED caused it, although one could argue that a more rapid expansion of high-powered money and more bank bailouts could have prevented the recession (Krugman, 2008).

In addition to Krugman (2008), various economists and financiers from different institutions (National Comission on the Causes of the Financial and Economic Crisis in the United States, 2022, pp. 353-354), ranging from Ben Bernanke to Jamie Dimon or John Mack, point out that the Great Depression of 1929 shares many similarities (both phenomenological and in terms of causes) with the 2008 recession, to the extent that it is often referred to as "The Great Recession of 2008" in different media. Therefore, they can be considered qualitatively equivalent, and we can analyze, in case the FED's approach was "aggressive enough", whether Friedman was right or not.

---

[4] The rate of surplus value, which expresses the degree of exploitation of labor power that is possible given a level of development of the productive forces, represented by the average-weighted organic composition of capital.
[5] For example, in his book *Free to Choose: A Personal Statement* or in its documental version.

As shown by Federal Reserve Bank of St. Louis (2023), the FED adopted an aggressive strategy of rapidly expanding the monetary base. However, this measure alone did not work, and it had to be accompanied by many other measures that played a more significant role (Yale School of Management, 2022), (Federal Communications Comission, 2023). These measures included the 2009 *American Recovery and Reinvestment Act*, which provided funds for infrastructure projects, social assistance programs, and tax cuts for workers, as well as programs for purchasing public debt, reducing short-term interest rates almost to zero, significant reductions in long-term interest rates, stimulating the economic activity of small businesses, improving people's current consumption capacity, among many other policies. Additionally, there were subsequent improvements in the labor market and the real estate market (where the crisis originated), which are natural after a crisis due to the cyclical behavior of macroeconomic fundamentals. As Krugman (2008) points out, "The Monetary History thesis has just taken a hit".

### 3.2. The Role of the Gold Price Today

To extend Marx's criticisms of Hume's quantity theory of money to the modern quantity theory of money, it is necessary to demonstrate the mechanism by which the US dollar (as the international currency par excellence –for now-) is directly linked to the price of gold. Therefore, it is undeniable that we need to provide an answer to the question: if prices determine the quantities of money in circulation and if the monetary unit is a sign of value (meaning that money is ultimately not fiduciary), what is the ultimate commodity that serves as the foundation of money since the fall of Bretton Woods? To address this, we must consider the following issues.

Firstly, as Greenspan noted, "Gold still represents the ultimate form of payment in the world. Fiat money in extremis is accepted by nobody. Gold is always accepted" (Lewis, 2007, p. 98).

Secondly, gold has an inverse relationship with the quantity of money in circulation and the value of the US dollar (Gilroy, 2014), (Reuters, 2015), (Vakil, 2023).

Thirdly, in 1981, after a decade of abandoning the gold standard (a decade characterized by high inflation, debt crises, and savings issues in the United States and the Western world in general), reputable economists, high-ranking officials, and former officials of US monetary policy, including Greenspan, saw stabilizing "the general level of prices and, by inference, the price of the gold bullion itself" as the only way to deal with the volatility of the time (Lewis, 2007, p. 295). Therefore, in 1982, Volcker formally abandoned the monetarist experiment and adopted policies aimed at "stabilizing the value of the dollar against gold and other commodities, controlling the strong fluctuations of the 1970s and early 1980s." This was supported by the Plaza Accord (1985) and the Louvre Accord (1987) (Lewis, 2019). This ushered in the period known as the *Great Moderation* (1982-2007), where Volcker's policies were continued by Greenspan, a period that was a "welcome period of relative calm after the volatility of the Great Inflation" (Hakkio, 2013). These claims are supported by Laffer (2010, p. 198), who had a meeting with the FED chairman[6] and, based on that meeting, along with Charles Kadlec, wrote an article titled *Has the FED Already Put Itself on a Price Rule?* published in The New York Times on October 28, 1962.

Then, after Bernanke took office, an economist who declared that his central influence was Friedman (Lewis, 2013, p. 5) the FED's policies diverged from those of previous administrations. Bernanke's tenure was accompanied by a decline in the dollar (Lewis, 2018). Subsequently, Yellen resumed the path of Volcker and Greenspan, and the same occurred in the first stage of Powell's tenure (Lewis, 2018).

However, gradually, Jerome Powell, the first investment banker in history to lead the Federal Reserve, has moved away from the gold standard that he seemed to adopt in his initial tenure at the Fed. He declared in mid-2019 that it would not be a good idea to stabilize the dollar's price around the price of gold because it would tie monetary policy to it, and for that reason, the FED could no longer maximize employment, in addition to the fact that no country uses it (Franck 2019). On this, two things must be pointed out.

---

[6] This statement essentially expresses that "Look, I have no idea what today's prices or today's inflation are. And we won't have that data for months. But I know exactly what commodity prices are going to be in the future" (Laffer, 2010, p. 198). The practical effect of this policy, as demonstrated, was the stability of the dollar against gold.

The first is that employment levels in periods when the Federal Reserve stabilized the dollar's price around the price of gold have shown at least the same macroeconomic stability for the US economy (if not more) as those in which this was not done. While it cannot be conclusively argued that the gold standard guarantees general macroeconomic stability, it can at least be concluded that stabilizing the dollar's price around the price of gold is not incompatible with maximizing employment.

The second is that since Bretton Woods, no country has used the gold standard because they have anchored their currencies to the dollar, and the dollar has been anchored (in different forms and intensities) to the gold standard. Therefore, Powell's second argument also does not seem to hold up.

**Figure 2**

*Trend-Cycle of Gold Closing Price (US$) by Empirical Mode Decomposition Method*

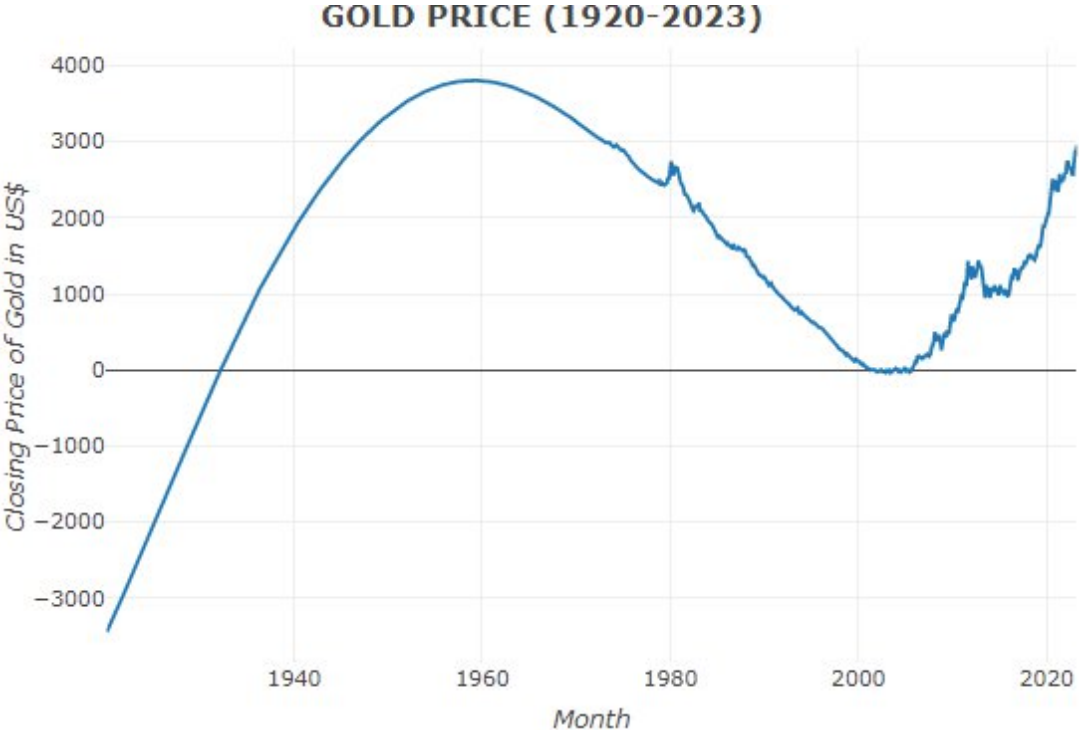

*Source:* Self-made in RStudio with data from the Bloomberg terminal.

As seen in Figure 2, starting with Bernanke's tenure in 2006, a period of increased volatility in the price of gold began, which was the most significant since the end of the Bretton

Woods era. After Yellen took office in 2014, the gold price tended to oscillate around an average during the first two years, followed by a marked linear trend in the next two years, except for a sharp drop in December 2016. Prior to the collapse of Bretton Woods, a smooth curve can be observed, while the curve starts to exhibit relatively mild volatility following the collapse of the dollar's convertibility into gold, which became significantly more pronounced with Bernanke's arrival. Of course, the decisions made by the FED chairman are just one element within the complex of material and objective conditions that explain the volatility of the gold price during their tenure.

From the above, it is possible to define the gold standard system in its most general form as one that "encompasses a large number of specific systems that could be created, each with its own idiosyncrasies, but all sharing the same goal of maintaining the currency's value at a specific gold parity" (Lewis, 2013, pp. 29-30). This implies two issues. On the one hand, the international financial system tends to be governed (especially when analyzing large periods of time) by an underlying gold standard, and on the other hand, this recurrence to the gold standard expresses that the international economic system naturally tends towards it due to the stability it provides.

Of course, an underlying gold standard is not as effective in ensuring stability as a rigorously established gold standard. The abandonment of the gold standard by the United States was something that a few economists managed to foresee, such as the Marxist economist Ernest Mandel. If the gold standard provides stability to the economic system, why is there a tendency among the real powers and economic policymakers in the United States to move away from it?

As Mandel (1968), points out, there is a specific set of reasons why the United States decided to abandon the gold standard. The gold standard implies that the U.S. government must implement contractionary economic policies, aiming to generate deflation when recessions coincide with deficits in the balance of payments. However, practicing contractionary policies in such scenarios can lead to economic crises of exceptional depth, as was the case with austerity policies (cutting government spending and increasing taxes) implemented by the Heinrich Bruning government in Germany between 1930 and 1932, which, in turn, can undermine democracy by causing the rise of fascist groups to power (Galofré-Vilà, et al 2020, pp. 2-6).

Additionally, the gold standard is unsustainable if there is no synchrony between investors increasing the demand for dollars (the rest of the world, primarily Europeans) and investments reducing the demand for dollars (Mandel, 1968). When European investors deposit their reserves in U.S. dollars in the short term, they are increasing the demand for dollars in the foreign exchange market. On the other hand, when U.S. investors invest their reserves in Europe for the long term, they are withdrawing dollars from the market and buying other currencies, reducing the demand for dollars in the foreign exchange market. This duality was precisely what characterized the years leading up to the collapse of Bretton Woods.

If this imbalance in the demand for dollars continues for a long enough period, it can lead to a monetary crisis by two paths. The first, if the supply of dollars in the market is insufficient to meet the demand from investors, despite initially causing an appreciation, persistent inability of the supply to generate demand can lead to a sharp drop in demand for the currency and, consequently, a sharp fall in the relative price of the currency. The second, if the demand for other currencies is greater than that for dollars, this situation can lead to a sudden and significant depreciation of the dollar, which can have adverse effects on the U.S. economy and the global economy (as a whole). This is what happened around the 1970s because, considering short and long-term actions and documents (bonds, treasury bills, and other securities issued by the government or companies), the demand for short-term dollars by Europeans was lower than the demand for long-term European currencies by Americans (Mandel, 1968). Speculation in the stock market should tend to exacerbate this duality in currency investments.

The attraction and repulsion effect of the gold standard within the capitalist system reveals that it inherently has both the need to converge to the gold standard and the need to distance itself from it, which expresses the fundamental dialectical contradiction of the international monetary system (Mandel, 1968):

> "The "dollar crisis" and the search for means of international payment independent both of gold and "currency reserves" (dollars and pounds sterling) reflect clear recognition on the part of big international capital of a contra-diction inherent in the present-day capitalist system; the contradiction between the dollar's role as an "international money," and its role as an instrument to assure the expansion of the American capitalist economy. To fulfill the first function, a stable money is needed. To fulfill the second function, a flexible money is necessary, *i.e.,* an unstable one. There's the rub."

The relationships expressed in this section between the price of gold, the dollar, and the quantity of money in circulation can be summarized in the following postulates:

1. If the price of gold rises, as a general rule, the quantity of money in circulation should decrease because fewer monetary units are needed to express the same sum of prices. However, some nuances must be considered. The increase in the price of gold can either raise or lower the price of the dollar (due to the inverse relationship between gold and the dollar) and, consequently, it can increase or decrease the quantity of money in circulation in the United States. This is because although in the United States, the same quantity of exchange values (commodities) now requires being expressed in a larger quantity of U.S. monetary units (the dollar), there will be corrective forces (both governmental and market-driven) that tend to stabilize the price of the dollar around the price of gold. Therefore, this inverse relationship between the dollar and gold will be a short-term trend, while in the long term, the relationship will be stable with periods where it is direct and periods where it is inverse; this periodic alternation is also linked, as Marx points out, to the dual role of gold (as a commodity and as a means of payment). On the other hand, the decrease in the price of the dollar implies a decrease in the quantity of money in circulation in countries anchored to the U.S. dollar because imports in these countries (whose prices are set in dollars) become cheaper as a result of needing a smaller quantity of domestic monetary units to represent the amount of US monetary units in which the commodity to be imported is valued according to the market (in these countries, the link between money and exchange values is both direct -with nominal GDP- and indirect -with the dollar-); an equivalent effect occurs when the price of gold falls because in these countries, there is no government policy to stabilize the currency's price relative to the price of gold.

2. If the price of gold falls, in general, the quantity of money in circulation should increase. This can either increase or decrease the price of the dollar and increase or decrease the quantity of money in circulation in the United States. Conversely, an increase in the price of the dollar implies an increase in the quantity of money in circulation in countries anchored to the U.S. dollar. All of this applies as long as all other factors are held constant.

**3.3. A Historically and Logically Grounded Mathematical Model of the Relationship Between Prices and Circulating Money**

All of the above applies as long as all other factors remain constant or as long as the described effect is quantitatively higher than other effects that counteract it (in which case it would be established as a trend). The most basic version of the above postulates can be expressed in mathematical terms as follows (Gómez Julián, 2016, p. 34):

$$Q_c = f\left(+\lambda_p, -\lambda_{gold}\right) : Q_m = \frac{\lambda_p}{\lambda_{gold}} \cdot \beta \qquad (1)$$

Where:

$Q_m$ = Circulating Money Supply (Monetary Aggregate M1).

$\lambda_p$ = Price of Goods (Nominal Gross Domestic Product).

$\lambda_{gold}$ = International Price of Gold.

$\beta$ = Transforming Coefficient representing the velocity of money circulation, therefore, it is equal to $\frac{1}{v}$, where $v$ is the velocity of money circulation. Here it is assumed to be an exogenous variable and equal to one.

The relationships between the variables in question can be expressed in terms of their rates of change as follows (Gómez, 2016, pp. 34-36):

1) If $\Delta\lambda_p < 0$ and $\Delta\lambda_{gold} = 0$ then $\Delta Q_m < 0$. So:

$$\frac{\partial Q_m}{\partial \lambda_p} = \frac{1}{\lambda_{gold}} \cdot \beta > 0$$

2) If $\Delta\lambda_p = 0$ and $\Delta\lambda_{gold} > 0$ then $\Delta Q_m < 0$. So:

$$\frac{\partial Q_m}{\partial \lambda_{gold}} = \frac{-1}{\lambda_{gold}^2}\cdot\beta < 0$$

3) If $\Delta\lambda_p < 0$ and $\Delta\lambda_{gold} < 0$ then $\Delta Q_m < 0$ if and only if $\Delta\lambda_p < 0 > \Delta\lambda_{gold} < 0$. So:

$$dQ_m = \frac{\partial Q_m}{\partial \lambda_p}d\lambda_p + \frac{\partial Q_m}{\partial \lambda_{gold}}d\lambda_{gold} = \frac{1}{\lambda_{gold}}\cdot\beta d\lambda_p - \frac{1}{\lambda_{gold}^2}\cdot\beta d\lambda_{gold}$$

Therefore, to obtain $\Delta Q_m < 0$, is necessary that:

$$\frac{1}{\lambda_{gold}}\cdot\beta d\lambda_p - \frac{1}{\lambda_{gold}^2}\cdot\beta d\lambda_{gold} < 0 \iff \frac{1}{\lambda_{gold}}\cdot\beta d\lambda_p < \frac{1}{\lambda_{gold}^2}\cdot\beta d\lambda_{gold}$$

4) If $\Delta\lambda_p > 0$ and $\Delta\lambda_{gold} > 0$ then $\Delta Q_m > 0$ if and only if $\Delta\lambda_p > 0 > \Delta\lambda_{gold} > 0$. Let's see the same case as the previous one, but in reverse:

$$dQ_m = \frac{\partial Q_m}{\partial \lambda_p}d\lambda_p + \frac{\partial Q_m}{\partial \lambda_{gold}}d\lambda_{gold} = \frac{1}{\lambda_{gold}}\cdot\beta d\lambda_p - \frac{1}{\lambda_{gold}^2}\cdot\beta d\lambda_{gold}$$

Therefore, to obtain $\Delta Q_m > 0$, is necessary that:

$$\frac{1}{\lambda_{gold}}\cdot\beta d\lambda_p - \frac{1}{\lambda_{gold}^2}\cdot\beta d\lambda_{gold} > 0 \iff \frac{1}{\lambda_{gold}}\cdot\beta d\lambda_p > \frac{1}{\lambda_{gold}^2}\cdot\beta d\lambda_{gold}$$

5) If $\Delta\lambda_p > 0$ and $\Delta\lambda_{gold} = 0$ then $\Delta Q_m > 0$, which is equivalent to point 1, confirming the direct relationship between the value of goods and the circulating money supply. This scenario would be tautological.

6) If $\Delta\lambda_p = 0$ and $\Delta\lambda_{gold} < 0$ then $\Delta Q_m > 0$, which is equivalent to point 2, confirming the inverse relationship between the value of gold and the circulating money supply. This scenario would be tautological.

7) If $\Delta\lambda_p > 0$ and $\Delta\lambda_{gold} < 0$ then $\Delta Q_m > 0$. Note that there is no ambiguity here. Given that the value of goods has a direct relationship with the circulating money supply (confirmed by point 5), and the value of gold has an inverse relationship with that supply (confirmed by point 6), the circulating money supply can only increase in this case. If the value of goods increases but the value of gold decreases, it means that more circulating money is needed than would be needed to express the value of goods if the latter did not change (this can also be seen easily with the total derivative of the circulating money supply).

8) Finally, if $\Delta \lambda_p > 0$ and $\Delta \lambda_{gold} < 0$ then $\Delta Q_m > 0$. Note that there is no ambiguity here either, as the value of goods has a direct relationship with the circulating money supply (confirmed by point 5), and the value of gold has an inverse relationship with that supply (confirmed by point 6). In this case, the circulating money supply can only increase. If the value of goods increases but the value of gold decreases, it means that more circulating money is needed than would be needed to express the value of goods if the latter did not change (this can also be seen easily with the total derivative of the circulating money supply).

Without loss of generality, it is possible to express the above relationships in terms of elasticities by applying a logarithmic transformation to both sides of the quantity equation to not alter the identity of its members and assuming $\beta = 1$:

$$ln(Q_m) = ln\left(\frac{\lambda_p}{\lambda_{gold}}\right) = ln(\lambda_p) - ln(\lambda_{gold}) \qquad (2)$$

This will allow, when constructing regression models, to interpret the model coefficients as elasticities, as well as smooth the relationship between the data, potentially improving the statistical results in certain tests.

A more general version of the case presented in equations 1 and 2, which encompasses the relationships established in the two hypotheses, is as follows:

$$Q_m = f\left(+\lambda_p, \pm \lambda_{gold}\right) \qquad (3)$$

The following section is dedicated to the experimental verification of the hypotheses presented in the preceding sections.

## 4. Econometric Analysis

### 4.1. Generalities

The econometric research of the studied phenomenon consists of several stages and analyzes four countries on a quarterly basis: the United States (1959-2022), Canada (1961-2022), the United Kingdom (1986-2022), and Brazil (1996-2022). All this data is taken from Gómez Julián (2024).

The United States is chosen because it is the most developed Western capitalist economy globally. Therefore, the economic laws derived from its study are, to a greater or lesser extent, applicable to other capitalist countries, regardless of their level of development. The U.S. economy, being the highest stage of development reached by Western capitalism, serves as a reflection of the future for all Western capitalist economies. The United Kingdom is selected for similar reasons as the United States, and there is also more data available for the UK compared to other candidates like Germany or France. Canada is chosen for similar reasons to the UK, as well as the fact that its model of Western capitalism is different from that of the UK, representing the welfare state variant. Finally, Brazil is selected as an emerging economy (developing country), and since this study has already been conducted for El Salvador (an underdeveloped country), verifying the local economic laws would have broad general implications. This makes it reasonable to assume that the results can be considered general economic laws of capitalist development.

Firstly, the research focuses on studying the determining direction of the feedback relationship between prices-money, gold-prices, and gold-money using univariate Bayesian linear regression models obtained through a Hamiltonian system of Markov Chain Monte Carlo and RESET tests for quadratic, cubic, and quadratic-cubic relationships robustified with Bayesian bootstrapping.

With these results, a generalized bivariate Bayesian linear model is formulated using different families and links that connect the three analyzed variables. Transformations are applied to these variables, such as spline regressions and empirical probability distribution fittings. The latter are done using the method of maximum goodness-of-fit, and the results of the distribution fit are chosen based on the Bayesian Information Criterion. Bayesian cross-validation is then applied using the "Leave-One-Out" (LOO) method.

Next, predictive performance metrics of the model (mean absolute error and root mean squared error), variance-explaining performance metrics (which are only applicable to models using Gaussian or binomial families), and other relevant model quality metrics are estimated.

Finally, different machine learning and deep learning models are explored, and the feasibility[7] of constructing ensembles between two or more candidates, specifically using the boosting mechanism, is assessed. These ensembles have the structure of a generalized Bayesian linear model with a Gaussian family and identity link. They also undergo the previously described cross-validation. The candidate models include a quantile random forest, a regularized Bayesian neural network, a support vector machine with a radial basis kernel, and a conditional inference random forest. The results are reported, allowing for comparison with the results of the preliminary generalized bayesian linear model (the one directly built with the studied variables).

Bayesian analysis is of an objective philosophical nature. The objective nature of Bayesian modeling is observed in that, in all cases, prior information, such as prior $R^2$ and prior intercept, is determined using empirical data from the dataset, rather than relying on subjective beliefs or assumptions about data behavior. This preliminary empirical information is not derived from raw data but is obtained through frequentist analyses on the data.

Approaching the problem from the standpoint of objective bayesianism allows for considering uncertainty about parameter estimates, which is epistemologically equivalent to incorporating epistemological doubt about parameter estimation. This not only has concrete ontological consequences but also reflects an ontological view of existence different from that implicit in the frequentist or subjective Bayesian approach. The methodology used, as will be explained in detail in the theoretical framework, allows for obtaining stationary distribu-tions of each of the estimated parameters, making it possible to analyze the long-term behavior of phenomena in terms of their first moment of probability.

---

[7] This feasibility is determined based on whether the ensemble provides greater robustness than any of the candidates analyzed individually.

The choice of generalized linear models is due to their ability to ensure that possible autocorrelation between prediction error terms or variable-variance does not affect parameter estimates. Regarding the choice of family and link in each generalized linear model directly constructed with the study variables, it was based on the results provided by predictive and explanatory metrics after running the model with different families and links, especially predictive metrics. This is because for models with families other than Gaussian or binomial, it is not possible to obtain the primary predictive metric for these models, the $R^2$. For ensembles, all were used with a Gaussian family and identity link because the higher computational cost of other families and links was not feasible. This was because the individual members of the ensemble showed outstanding predictive and explanatory performance, and the resulting ensembles, even when not superior to one or more of their individual components, also had outstanding predictive and explanatory performance.

**4.2. The Case of the United States (1959-2022)**

**Table 1**

*Results of Bayesian Simple Linear Regression Analysis*

| Variable 1 | Variable 2 | Best Model | Criterions |
|---|---|---|---|
| $log(M1)$ | $log(Prices)$ | Undecidable | log-fit ratio and ELPD-LOO |
| $log(Gold)$ | $log(Prices)$ | $log(Gold) = f(log(Prices))$ | log-fit ratio and ELPD-LOO |
| $log(M1)$ | $log(Gold)$ | $log(M1) = f(log(Gold))$ | log-fit ratio and ELPD-LOO |

**Table 2**

*Mean p-value from Posterior Bootstrap Distribution of RESET Tests*

| Functional Form | RESET of Squares | RESET of Cubes | RESET of Squares and Cubes |
|---|---|---|---|
| $\log(M1) = f(\log(Prices))$ | 0 | 0 | 0 |
| $\log(Prices) = f(\log(M1))$ | 0 | 0 | 0 |
| $\log(Prices) = f(\log(Oro))$ | 0.1025 | 0.091 | 0.1057 |
| $\log(Gold) = f(\log(Prices))$ | 0.027 | 0.051 | 0 |
| $\log(Gold) = f(\log(M1))$ | 0 | 0 | 0 |
| $\log(M1) = f(\log(Gold))$ | 0 | 0 | 0 |

**Table 3**

*Results of the Empirical Distribution Fitting by the Maximum Goodness of Fit Method*

| Variable | Distributions | BIC | Parameter 1 | Parameter 2 |
|---|---|---|---|---|
| log(M1) | Log-normal | 809.2976 | Location = 1.88 | Scale = 0.19 |
|  | Gamma | 811.5362 | Location = 28.60 | Rate = 4.30 |
| log(Prices) | Weibull | 819.6427 | Form = 6.25 | Scale = 7.65 |
|  | Log-normal | 854.0632 | Location = 1.96 | Scale = 0.20 |
| Gold | Weibull | 3730.965 | Form = 0.89 | Scale = 515.76 |
|  | Log-normal | 3748.047 | Location = 5.78 | Scale = 1.29 |

The previous econometric results are consistent with the analyses conducted in the preceding sections. These results, along with the contrast between the outcomes of applying BGLM with different families, links, and transformations on the variables, led us to formulate the following BGLM with a Gamma family, logarithmic link, and the variable 'Gold' transformed into a natural cubic spline (ns) with five degrees of freedom (df) and five basis functions of the form $ns(ORO, df = 5)_i$ con $i = 1, ..., 5$:

$$log(M1) \approx 0.97 + 0.13 \cdot log(Prices) - 0.35 \cdot ns(Gold, df = 5)_1 - 0.01 \cdot ns(Gold, df = 5)_2 - 0.04 ns(Gold, df = 5)_3 + 0.09 \cdot ns(Gold, df = 5)_4 + 0.21 \cdot ns(Gold, df = 5)_5 \qquad (4)$$

The model was chosen based on MAE (Mean Absolute Error), RMSE (Root Mean Squared Error)[8], ELPD-LOO[9] (Expected Log Pointwise Predictive Density -ELPD- from Leave-One-Out Cross-Validation), P-LOO (effective number of LOO parameters), LOO-IC (LOO Information Criterion), SE-MC of ELPD-LOO (Monte Carlo Standard Error of ELPD-LOO), PSIS-LOO (Pareto Importance Sampling of LOO), and the GVIF (Generalized Variance Inflation Factor) corrected for degrees of freedom. The coefficient of determination $R^2$ was not considered because it can only be calculated for GLM's with Gaussian or binomial families. The relevant statistics of the model expressed in equation 4 are presented below.

**Table 4**

*Summary of Model Performance Metrics*

| Statistic | Value |
| --- | --- |
| Range of log(M1) | [4.94, 9.94] |
| Prior mean of log(M1) | 6.64 |
| Posterior predictive distribution mean of log(M1) | 6.6 |
| Median of M1 | 6.71 |
| MAE | 0.12 (2.43% of the log(M1) minimum) |
| RMSE | 0.21 (4.25% of the log(M1) minimum) |
| ELPD-LOO | -122.5 |
| P-LOO | 1.3 |
| LOO-IC | 245.1 |

---

[8] Since MAE and RMSE are comparative metrics between models, and here only one functional form is modeled after discarding other candidates in the preliminary analysis, these metrics will be expressed as a proportion of the minimum value of the predicted response variable to show the maximum estimation error incurred by the model. This allows these metrics to make sense without the need to compare different models.

[9] As stated in (Vehtari, Gelman & Gabry 2016, pp. 19-21).

| | |
|---|---|
| k values of PSIS-LOOO | < 0.05 |
| SE-MC of ELPD-LOO | 0 |
| GVIF corrected by df | $\log(NGDP) = 3.21$, $ns(Oro, df = 5) = 1.26$ |

The resulting multivariate model is consistent in terms of the signs of its coefficients with the general case discussed in the preceding sections and expressed mathematically in equation 3. If we observe the coefficients related to gold, their relationship with the money supply varies by segment: sometimes it is inverse, and sometimes it is direct. The direct relationship between the money supply and prices is also confirmed.

**Figure 3**

*Graphical Predictive Performance of the Multivariate Model*

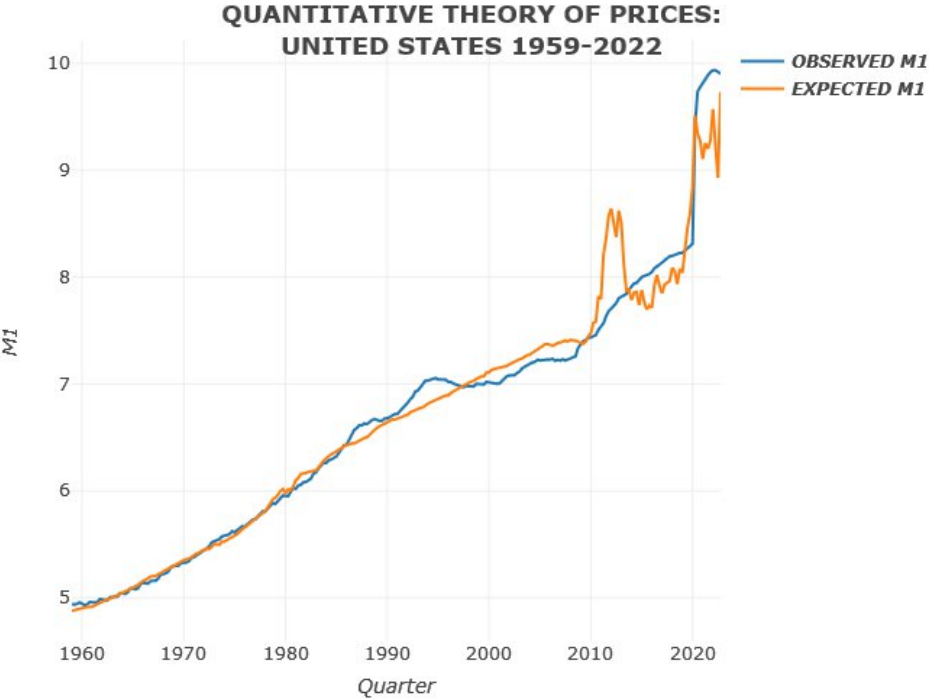

Additionally, it is possible to explore the possibility of improving the performance metrics of the multivariate model by constructing an ensemble of models, specifically by boosting two or more models within a Bayesian Generalized Linear Model (BGLM) with a Gaussian family and identity link. The ensemble BGLM, consisting of a Bayesian Regularized Neu-ral Network (BRNN) and a Quantile Random Forest (QRF), was constructed after individually evaluating and pairwise ensemble testing a Conditional Inference Random Forest, a Quantile Random Forest, a Regularized Bayesian Neural Network, and a Support Vector Machine with a Radial Basis Function kernel. Each of these models was selected from possible candidates generated by a specific parameter grid for each of them. The metrics used to choose the best model, based on a repeated cross-validation process with 10 folds and 100 repetitions using a 20% test partition, included the $R^2$, the Akaike Information Criterion (AIC), the Mean Absolute Error (MAE), the Root Mean Square Error (RMSE), and residual deviance vs. null deviance. The chosen metrics were based on test results. The relevant statistics are presented below.

**Table 5**

*Analysis of Ensemble Coefficients (with test data)*

| Coefficient | Value | Standard Error | t-value | Pr(>|t|) |
|---|---|---|---|---|
| Intercept | -0.031108 | 0.005865 | -5.304 | 1.14e-07 *** |
| BRNN | 0.410388 | 0.00628 | 65.345 | < 2e-16 *** |
| QRF | 0.594979 | 0.006307 | 94.339 | < 2e-16 *** |
| Significance codes: 0 '***' 0.001 '**' 0.01 '*' 0.05 '.' 0.1 ' ' 1 | | | | |
| Dispersion parameter of the Gaussian family set to 0.0218805 | | | | |
| Null Desviance: 29749.09 with 20799 df | | | | |
| Residual Deviance: 455.05 with 20797 df | | | | |
| AIC: -20468 | | | | |
| Number of iterations by Fisher Scoring method: 3 | | | | |

**Table 6**

*Repeated Cross-Validation Results for the Ensemble*

| Statistic | Value |
|---|---|
| Range of M1 (training) | [23.06, 28.12] |
| Prior mean of M1 (training) | 25.47 |
| Median of M1 (training) | 25.60 |
| MAE (training) | 0.06 (0.26% of the log(M1) minimum) |
| RMSE (training) | 0.15 (0.65% of the log(M1) minimum) |
| $R^2$ (training) | 0.985 |
| Range of M1 (test) | [23.05, 28.08] |
| Prior mean of M1 (test) | 25.52 |
| Median of M1 (test) | 25.53 |
| MAE (test) | 0.08 (0.34% of the log(M1) minimum) |
| RMSE (test) | 0.25 (1.06% of the log(M1) minimum) |

**Figura 4**

*Graphic Predictive Performance of the Ensamble*

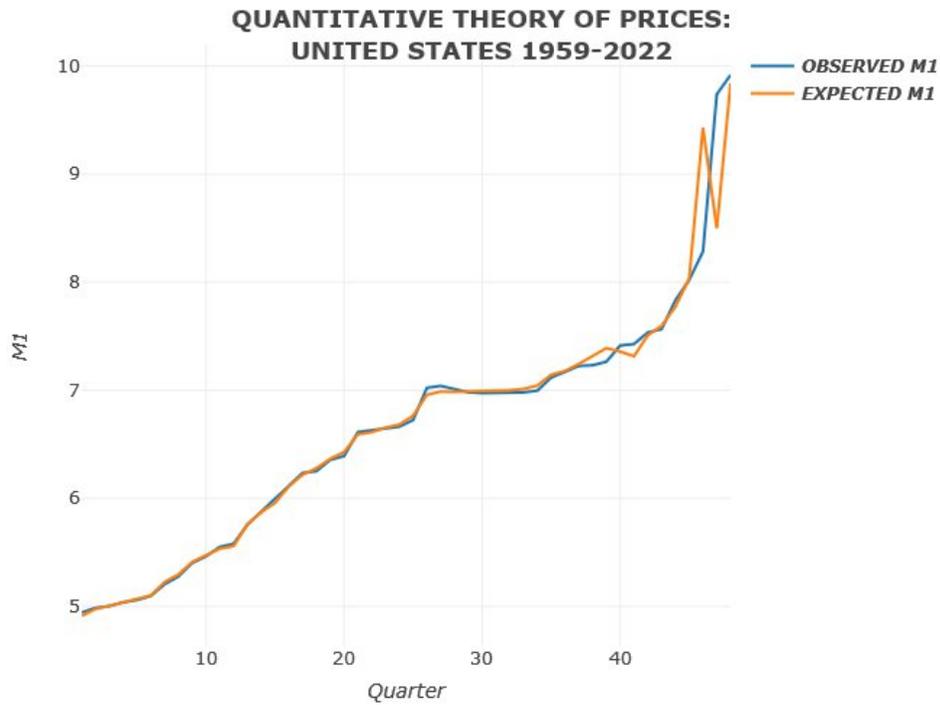

### 4.3. The Case of Canada (1961-2022)

**Table 7**

*Results of Bayesian Simple Linear Regression Analysis*

| Variable 1 | Variable 2 | Best Model | Criterion |
|---|---|---|---|
| $log(M1)$ | $log(Prices)$ | $log(M1) = f(log(Prices))$ | ELPD-LOO |
| $log(Gold)$ | $log(Prices)$ | $log(Gold) = f(log(Prices))$ | ELPD-LOO |
| $log(M1)$ | $log(Gold)$ | $log(M1) = f(log(Gold))$ | ELPD-LOO |

**Table 8**

*Mean p-value from the Posterior Bootstrap Distribution of RESET Tests*

| Functional Form | RESET of Squares | RESET of Cubes | RESET of Squares and Cubes |
|---|---|---|---|
| $log(M1) = f(log(Prices))$ | 0 | 0 | 0 |
| $log(Prices) = f(log(M1))$ | 0 | 0 | 0 |
| $log(Prices) = f(log(Gold))$ | 0.092 | 0.074 | 0.026 |
| $log(Gold) = f(log(Prices))$ | 0.045 | 0.0562 | 0.0007 |
| $log(Gold) = f(log(M1))$ | 0 | 0 | 0 |
| $log(M1) = f(log(Gold))$ | 0 | 0 | 0 |

**Table 9**

*Results of the Empirical Distribution Fitting by the Maximum Goodness of Fit Method*

| Variable | Distributions | BIC | Parameter 1 | Parameter 2 |
|---|---|---|---|---|
| log(M1) | Normal | 904.8703 | Location = 25.5 | Scale = 1.69 |
| | Gamma | 905.3589 | Location = 227.96 | Rate = 8.93 |
| log(Prices) | Weibull | 789.4102 | Form = 10.85 | Scale = 12.39 |
| | Normal | 817.3348 | Location = 11.91 | Scale = 1.32 |
| log(Gold) | Weibull | 808.6559 | Form = 5.88 | Scale = 6.47 |

|  | Normal | 837.2897 | Location = 6.06 | Scale = 1.20 |
|---|---|---|---|---|

The previous econometric results are consistent with the analyses conducted in the preceding sections. These results, in combination with the contrast between the outcomes of applying BGLM with different families, links, and transformations on the variables, led us to propose the following BGLM with a Gamma family, logarithmic link, and the variable 'Gold' transformed into a $Weibull\,(shape = 5.883053, scale = 6.473397)$ random variable.

$$log\,(M1) \approx 2.695 + 0.045 \cdot log\,(PIBN) - 0.00001982922 \cdot weibull\,(shape = 5.88, scale = 6.47) \quad (5)$$

The model, among all the tested candidates, was chosen based on the same metrics used in the case of the United States. The relevant statistics of the model expressed in equation 5 are presented below.

**Table 10**

*Summary of Model Performance Metrics*

| Statistic | Value |
|---|---|
| Range of log(M1) | [23.05, 28.12] |
| Prior mean of log(M1) | 25.48 |
| Posterior Predictive distribution mean of log(M1) | 25.4 |
| Median of log(M1) | 25.59 |
| MAE | 0.23 (1% of the log(M1) minimum) |
| RMSE | 0.28 (1.21% of the log(M1) minimum) |
| ELPD-LOO | -436.5 |
| P-LOO | 0.6 |
| LOO-IC | 873 |
| k values of PSIS-LOOO | < 0.05 |
| SE-MC of ELPD-LOO | 0 |
| GVIF | $log\,(NGDP) = 1$, $Gold \sim Weibull = 1$ |

**Figure 5**

*Graphical Predictive Performance of the Multivariate Model*

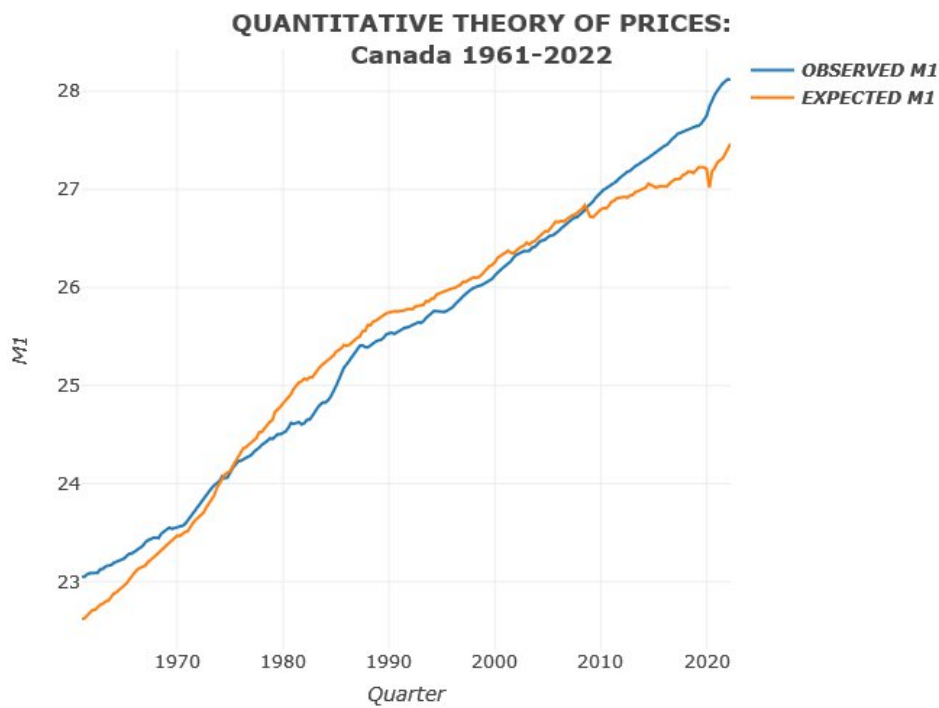

The resulting multivariate model is congruent in terms of the signs of its coefficients with the basic case expressed in equation 1 since, if you observe the coefficients of gold, their relationship with the circulating money varies depending on the segment: sometimes it is inverse, and sometimes it is direct. The direct relationship between the circulation and prices is also confirmed. In fact, equations 2 and 5 are theoretically equal since the constructed Weibull random variable expresses the best-fitting probability distribution of the variable 'Gold'.

Unlike the US case, where it was possible to improve the performance metrics of the original multivariable model by constructing a model ensemble, for the Canadian case, while ML models were constructed that improved the model in equation 5, the ensembles did not improve the individual performance of these models. When evaluated with the same models and using the same methodology as in the US case, a Quantile Random Forest (QRF) was the model that showed the best performance in training and testing under the same criteria as in the previously studied case.

**Table 11**

*Results of Repeated Cross-Validation for QRF*

| Statistic | Value |
|---|---|
| Range of M1 (training) | [23.06 28.12] |
| Mean of M1 (training) | 25.47 |
| Median of M1 (training) | 25.60 |
| MAE (training) | 0.04 (0.15% of the log(M1) minimum) |
| RMSE (training) | 0.06 (0.26% of the log(M1) minimum) |
| $R^2$ (training) | 0.9983 |
| Range of M1 (test) | [23.05, 28.08] |
| Mean of M1 (test) | 25.52 |
| Median of M1 (test) | 25.53 |
| MAE (test) | 0.04 (0.17% of the log(M1) minimum) |
| RMSE (test) | 0.08 (0.35% of the log(M1) minimum) |

**Figure 6**

*Graphical Predictive Performance of the QRF*

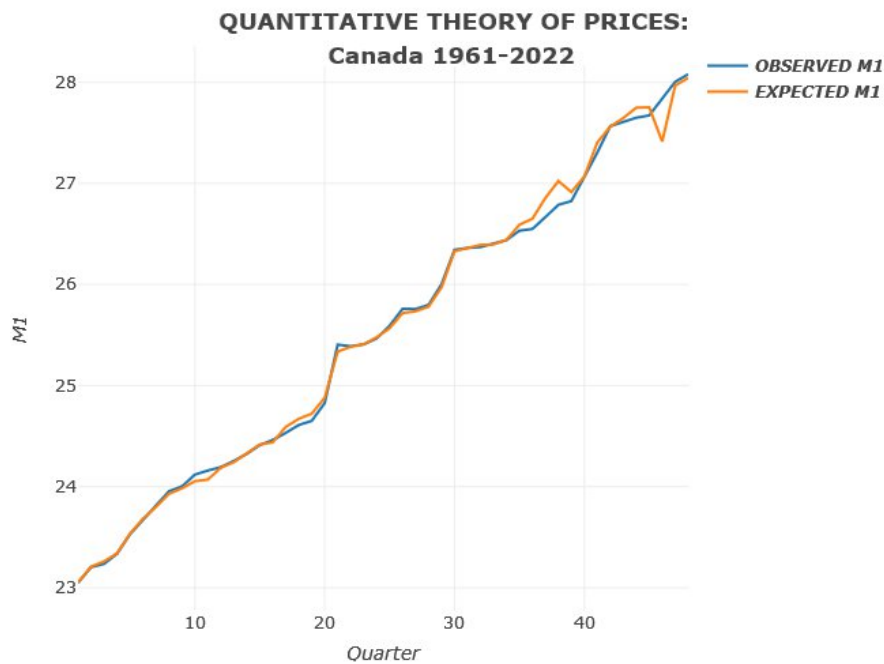

## 4.4. The Case of the United Kingdom (1986-2022)

**Table 12**

*Results of Bayesian Simple Linear Regression Analysis*

| Variable 1 | Variable 2 | Best Model | Criterion |
|---|---|---|---|
| $log(M1)$ | $log(Prices)$ | $log(Prices) = f(log(M1))$ | ELPD-LOO |
| $log(Gold)$ | $log(Prices)$ | $log(Gold) = f(log(Gold))$ | ELPD-LOO |
| $log(M1)$ | $log(Gold)$ | $log(M1) = f(log(Gold))$ | ELPD-LOO |

**Table 13**

*Average p-value from the Posterior Bootstrap Distribution of RESET Tests*

| Functional Form | RESET of Squares | RESET of Cubes | RESET of Squares and Cubes |
|---|---|---|---|
| $log(M1) = f(log(Prices))$ | 0 | 0 | 0 |
| $log(Prices) = f(log(M1))$ | 0 | 0 | 0 |
| $log(Prices) = f(log(Gold))$ | 0.21 | 0.26 | 0.026 |
| $log(Gold) = f(log(Prices))$ | 0 | 0 | 0 |
| $log(Gold) = f(log(M1))$ | 0 | 0 | 0 |
| $log(M1) = f(log(Gold))$ | 0.20 | 0.24 | 0 |

**Table 14**

*Results of the Empirical Distribution Fitting by the Maximum Goodness of Fit Method*

| Variable | Distributions | BIC | Parameter 1 | Parameter 2 |
|---|---|---|---|---|
| log(M1) | Weibull | 400.2271 | Location = 29.59 | Scale = 27.55 |
|  | Normal | 405.1268 | Location = 27.14 | Scale = 1.08 |
| log(Prices) | Weibull | 184.3187 | Form = 61.95 | Scale = 26.69 |
|  | Normal | 192.2247 | Location = 26.50 | Scale = 0.50 |
| Gold | Log-normal | 2086.895 | Form = 5.96 | Scale = 0.93 |
|  | Gamma | 2102.225 | Location = 1.29 | Rate = 0.002 |

The previous econometric results are consistent with the analyses conducted in the preceding sections. Similar to the case of the United States, the aforementioned results, in conjunction with the comparison between the outcomes of applying BGLM with different families, links, and transformations on the variables, led us to propose the following BGLM with a Gamma family, logarithmic link, and the variable 'Gold' transformed into a natural cubic spline (ns) with five degrees of freedom and five basis functions of the form $ns(Gold, df = 5)_i$ with $i = 1, ..., 5$:

$$log(M1) \approx 1.42 + 0.071 \cdot log(NGDP) + 0.0001 \cdot ns(Gold, df = 5)_1 + 0.012 \cdot ns(Gold, df = 5)_2 - 0.001 ns(Gold, df = 5)_3 + 0.006 \cdot ns(Gold, df = 5)_4 + 0.01 \cdot ns(Gold, df = 5)_5 \quad (6)$$

The resulting multivariable model is congruent in terms of the signs of its coefficients with the general case outlined in the preceding sections and mathematically expressed in formula 3, as the coefficients of gold exhibit varying relationships with the circulating money, sometimes inverse and sometimes direct. The direct relationship between the circulating money and prices is also confirmed.

The model was chosen among all the candidates based on the same metrics used in the previous two cases. The relevant statistics of the model expressed in equation 6 are presented below.

**Table 15**

*Summary of Model Performance Metrics*

| Estadístico | Valor |
|---|---|
| Range of log(M1) | [25.20, 28.54] |
| Prior mean of log(M1) | 27.10 |
| Predictive posterior distribution mean of log(M1) | 27 |
| Median of log(M1) | 27.22 |
| MAE | 0.06 (0.24% of the log(M1) minimum) |
| RMSE | 0.07 (0.28% of the log(M1) minimum) |
| ELPD-LOO | -308.1 |
| P-LOO | 0.8 |
| LOO-IC | 616.2 |
| SE-MC of ELPD-LOO | 0 |
| k values of PSIS-LOOO | 0.05 |
| GVIF corrected by df | $log(PIBN) = 1.84, ns(ORO, df = 5) = 1.13$ |

**Figure 7**

*Graphical Predictive Performance of the Multivariate Model*

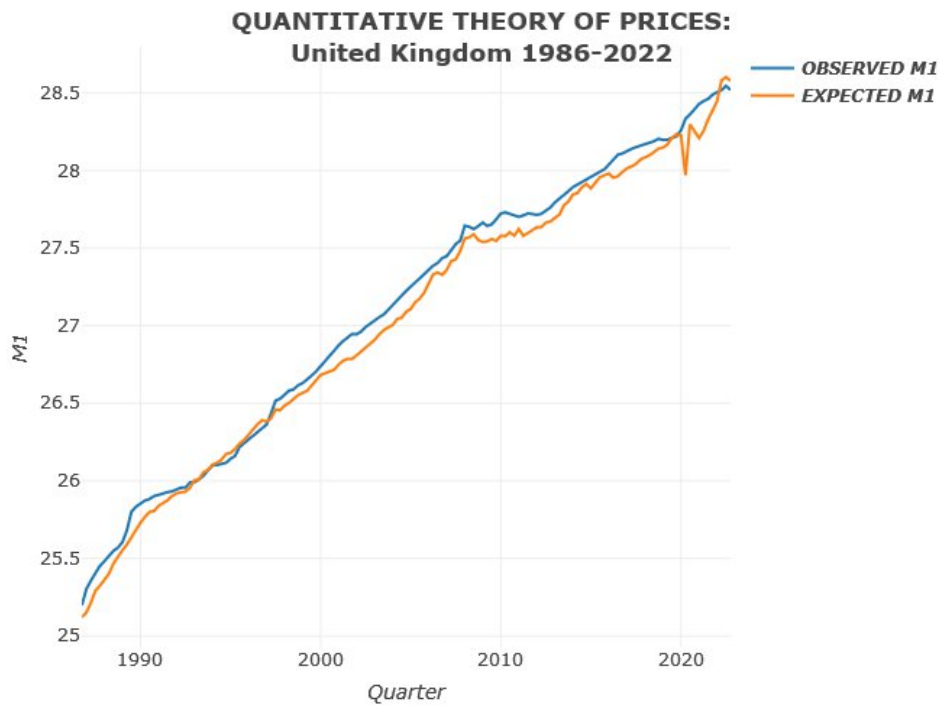

Unlike the U.S. case, where it was possible to improve the performance metrics of the original multivariable model by constructing an ensemble of models, for the case of the United Kingdom, although ML models were built that improved the model in equation 6, the ensembles did not improve the individual performance of these models. Evaluated using the same models and methodology as in the previous cases, a QRF was the model that demonstrated the best performance in training and testing under the same criteria as the previously studied case.

**Table 16**

*Repeated Cross-Validation Results of the QRF*

| Statistic | Value |
|---|---|
| Range of M1 (training) | [25.20, 28.50] |
| Media M1 (training) | 27.10 |
| Mediana de M1 (training) | 27.22 |
| MAE (training) | 0.05 (0.20% of the log(M1) minimum) |
| RMSE (training) | 0.08 (0.32% of the log(M1) minimum) |
| $R^2$ (training) | 0.993 |
| Range of M1 (test) | [25.20, 28.50] |
| Mean of M1 (test) | 27.07 |
| Median of M1 (test) | 227.21 |
| MAE (test) | 0.04 (0.16% of the log(M1) minimum) |
| RMSE (test) | ≈ 0 |

**Figure 8**

*Graphical Predictive Performance of the QRF*

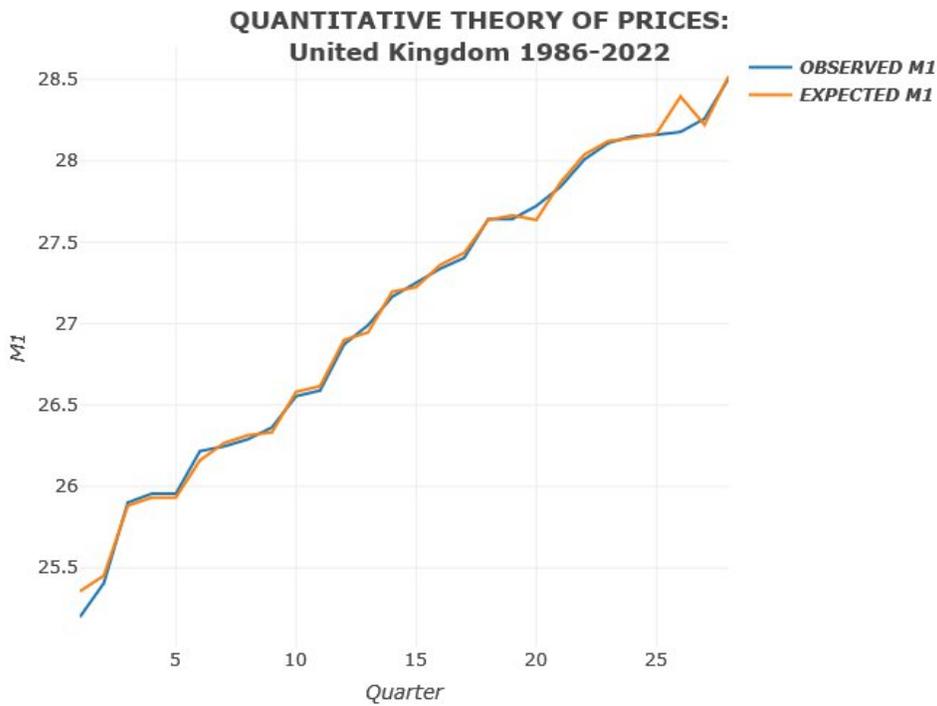

## 4.5. The Case of Brazil (1996-2022)

**Table 17**

*Results of Bayesian Simple Linear Regression Analysis*

| Variable 1 | Variable 2 | Best Model | Criterions |
|---|---|---|---|
| $log(M1)$ | $log(Prices)$ | $log(Prices) = f(log(M1))$ | log-fit ratio and ELPD-LOO |
| $log(Gold)$ | $log(Prices)$ | $log(Prices) = f(log(Gold))$ | log-fit ratio and ELPD-LOO |
| $log(M1)$ | $log(Gold)$ | $log(M1) = f(log(Gold))$ | log-fit ratio and ELPD-LOO |

Table 18

*Mean p-Value from the Bootstrap Posterior Distribution of RESET Tests*

| Functional Form | RESET of Squares | RESET of Cubes | RESET of Squares and Cubes |
|---|---|---|---|
| $\log(M1) = f(\log(Prices))$ | 0 | 0 | 0 |
| $\log(Prices) = f(\log(M1))$ | 0 | 0 | 0 |
| $\log(Prices) = f(\log(Gold))$ | 0.00015 | 0 | 0 |
| $\log(Gold) = f(\log(Prices))$ | 0.0074 | 0.006 | 0 |
| $\log(Gold) = f(\log(M1))$ | 0 | 0 | 0 |
| $\log(M1) = f(\log(Gold))$ | 0 | 0 | 0 |

Table 19

*Results of the Empirical Distribution Fittingt by the Maximum Goodness of Fit Method*

| Variable | Distributions | BIC | Parameter 1 | Parameter 2 |
|---|---|---|---|---|
| log(M1) | Weibull | 281.8004 | Form = 32.26 | Scale = 26.32 |
|  | Normal | 291.7062 | Location = 25.96 | Scale = 0.97 |
| log(Prices) | Weibull | 259.7543 | Form = 17.63 | Scale = 13.95 |
|  | Normal | 263.3563 | Location = 13.60 | Scale = 0.91 |
| Gold | Log-normal | 1939.814 | Location = 2288.59 | Scale = 2094.449 |
|  | Weibull | 1944.464 | Form = 1.04 | Scale = 2486.63 |

The previous econometric results are consistent with the analyses conducted in the preceding sections. These results, together with the comparison of results obtained by applying BGLM with different families, links, and transformations on the variables, led us to propose the following BGLM with a Gaussian family, logarithmic link, and the 'Gold' variable transformed into a natural cubic spline (ns) with five degrees of freedom and five basis functions of the form $ns(ORO, df = 5)_i$ with $i = 1,...,5$:

$$\log(M1) \approx 2.6 + 0.05 \cdot \log(NGDP) + 0.01 \cdot ns(Gold, df = 5)_1 + 0.002 \cdot ns(Gold, df = 5)_2 - 0.02 \cdot ns(Gold, df = 5)_3 + 0.008 \cdot ns(Gold, df = 5)_4 + 0.0006 \cdot ns(Gold, df = 5)_5 \quad (7)$$

The model, among all the candidates tested, was chosen based on the same metrics used in the previous two cases. The relevant statistics of the model expressed in equation 7 are presented below.

**Table 20**

*Summary of Model Performance Metrics*

| Statistic | Value |
| --- | --- |
| Range of log(M1) | [23.86, 27.21] |
| Prior mean of log(M1) | 25.88 |
| Predictive posterior distribution mean of log(M1) | 25.9 |
| Median of M1 | 26.16 |
| MAE | 0.07 (0.29% of the log(M1) minimum) |
| RMSE | 0.21 (0.42% of the log(M1) minimum) |
| ELPD-LOO | 88.3 |
| P-LOO | 9.5 |
| LOO-IC | -176.5 |
| k values of PSIS-LOOO | < 0.05 |
| SE-MC of ELPD-LOO | 0 |
| GVIF corrected by df | $\log(NGDP) = 7.04, ns(Gold, df = 5) = 1.48$ |

The resulting multivariable model is consistent in terms of the signs of its coefficients with the general case presented in the preceding sections and mathematically expressed in formula 3, as if we observe the coefficients of gold, their relationship with the circulating money varies depending on the range: sometimes it is inverse, and sometimes it is direct. The direct relationship between the circulating money and prices is also confirmed.

**Figure 9**

*Graphical Predictive Performance of the Multivariate Model*

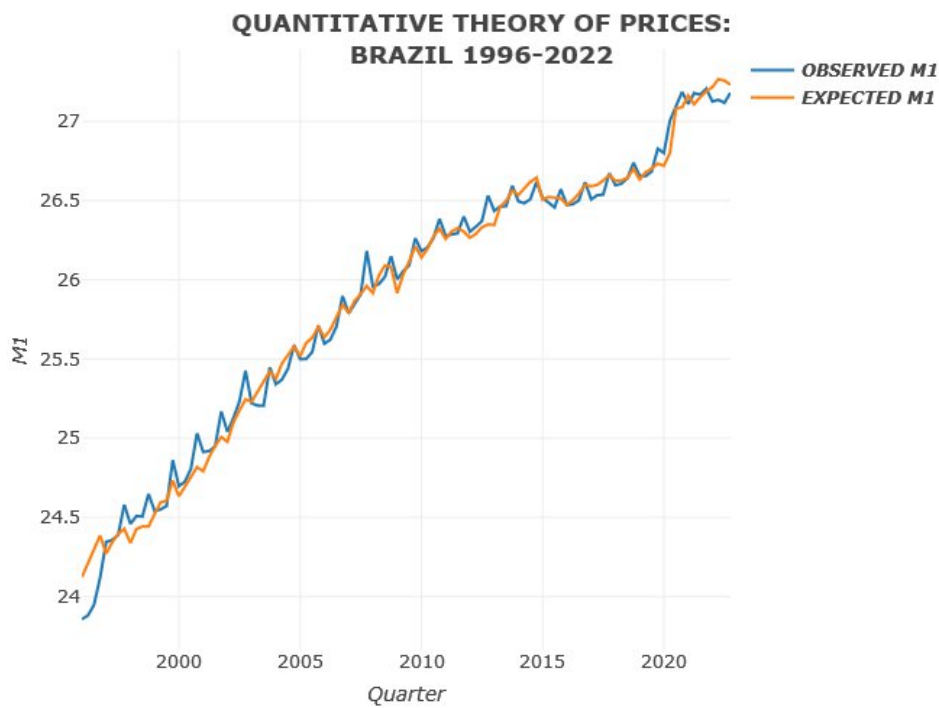

In contrast to the U.S. case, where it was possible to improve the performance metrics of the original multivariable model by constructing an ensemble of models, for the Brazilian case (as well as for Canada and the UK), while machine learning models were constructed that improved the model of equation 6, ensembles did not improve the individual performance of these models. Evaluated using the same models and methodology as in the previous cases, a Radial Basis Function Support Vector Machine (SVMRadial) was the model that demonstrated the best performance in training and testing under the same criteria as the previously studied case.

**Table 21**

*Repeated Cross-Validation Results for the SVMRadial*

| Statistic | Value |
|---|---|
| Range of log(M1) (training) | [5.82, 27.21] |
| Media log(M1) (training) | 25.90 |
| Mediana de log(M1) (training) | 26.16 |
| MAE (training) | 0.013 (1.27% of the log(M1) minimum) |
| RMSE (training) | 0.017 (1.67% of the log(M1) minimum) |
| $R^2$ (training) | 0.991 |
| Range of log(M1) (test) | [5.86, 27.13] |
| Prior mean of log(M1) (test) | 25.81 |
| Prior median log(M1) (test) | 25.90 |
| MAE (test) | 0.012 (1.20% of the log(M1) minimum) |
| RMSE (test) | 0.016 (1.62% of the log(M1) minimum) |

**Figure 10**

*Graphical Predictive Performance of the SVMRadial*

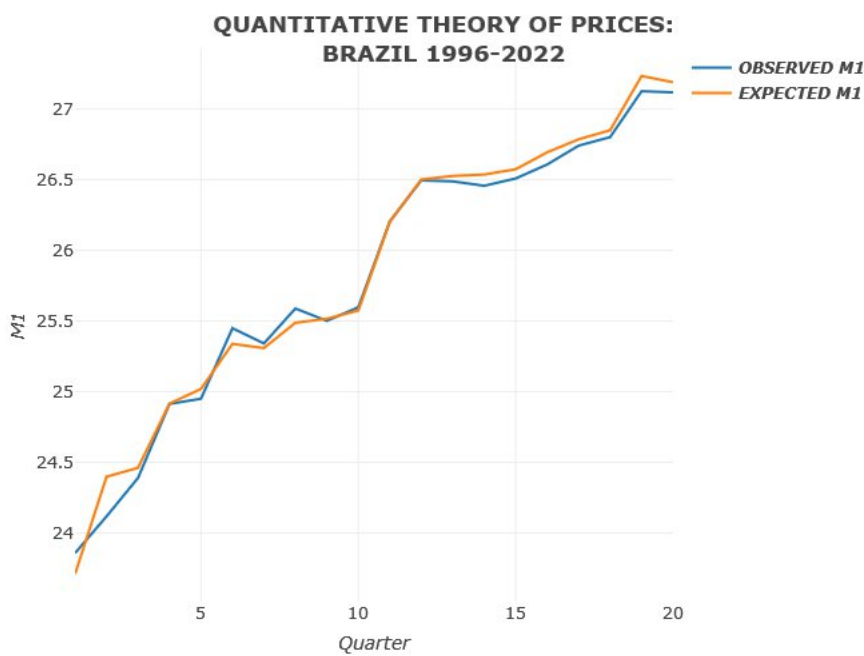

## 5. Conclusions

After analyzing the maximum sample size possible for four various countries, as well as capitalist economies with different degrees of development, where the sample size in some countries exceeded the length of a "Kondratieff wave"[10] (63 years for the United States, 61 years for Canada, 36 years for the United Kingdom, and 26 years for Brazil), the results of this research show that money is not neutral, neither in the short term nor in the long term. Because the results come from a theoretical framework whose fundamental postulates are antagonistic to that of the neoclassical framework, it raises a new question[11]: What is the cause of this non-neutrality?

What is explained by Marx (2008, 149-152) shows how the effects of the non-neutrality of money are transmitted to prices both in the short and long term, as well as the reason why this happens. Thus, non-neutrality is not caused by the circulating money determining prices, but by the mediating relationship it establishes between prices and the real foundation of money (gold and silver at that time) that determines the quantity of circulating money. This implies that this relationship has dynamic feedback (feedback over time) and taken this together with the econometric evidence showing that this feedback is nonlinear, it implies that the variables constitute a complex system in the sense of chaos theory (Willy, Neugebauer & Gerngroß, 2003, 19), (Raducha & San Miguel, 2020, 1), (Henderson-Sellers & McGuffie, 2012, 392).

However, since, as defined in the preceding sections, there has been no strict gold standard since the fall of Bretton Woods, but a loose one, the previous answers raise two new questions:

1. Now that there is no strict gold standard and what the FED does is stabilize the price of the dollar around gold (loose gold standard), how is the feedback relationship over time between prices and gold with the money supply established, and how is it ultimately that M1 is still determined by prices and gold?

---

[10] This is how orthodox economics knows the long waves of capitalist dynamics.
[11] Because, to the best of our knowledge, the available Marxist theory has not examined money from the perspective of neutrality or non-neutrality in an explicit manner, seeking to answer, from its own foundations, the questions that have been erroneously believed to be answered by neoclassical theory.

The relationship between M1 and prices, as well as its mechanism, remains in general terms as in the times of Hume and Marx. The aspect to be elucidated is then the relationship between M1 and the price of gold, given the fall of Bretton Woods. If the FED stabilizes the price of the dollar around the price of gold, as argued in the preceding sections, this implies that it will have to control the quantity of currency in the economy directly or indirectly through expansive/contractive economic policies. Thus, the relationship between M1 and the price of gold becomes complex, becoming a mediating relationship where the mediators are the instruments of economic policy of the U.S. government that serve to stabilize the price of the dollar around the price of gold. The specifics of how the feedback relationship over time between prices and gold with the money supply, involving economic policy instruments as latent variables, is a matter that requires further research.

Therefore, the limits of M1 variations are determined by the limits of price adjustments to variations in the exchange value of the subset of gold that serves to measure the values of commodities (what would this subset of gold that differs from the rest of the gold in modern times be, analogous to the subset mentioned by Marx in Hume's time?)

2. If money is never neutral, what are the quantitative and temporal limits of its effects?

The quantitative limit of the effects or non-neutrality of M1 is given by the fact that if the quantity of value tokens in circulation decreases or increases below or above its necessary level, a coercive correction will occur in M1 through an adjustment of commodity prices. This occurs because commodity prices determine the circulating money, so by adjusting the former, a correction to the latter is made. The time limit is determined by the time it takes for the mentioned adjustment to occur. This could open the possibility of new insights into phenomena that are traditionally attributed to having a monetary nature (for example, the 'liquidity trap').

In terms of economic policy, the above theoretically justifies an obvious point: that the best way to control prices is to do so directly. However, it also theoretically justifies a commonly effective practice in economic policy: that it is possible to do so by contracting M1 (directly or indirectly, depending on the dominant monetary regime for a given economic system). Thus, our approach is not only consistent with facts but also provides a broader and deeper explanation of them.